\documentclass[sigconf]{acmart}
\AtBeginDocument{%
  }

\copyrightyear{2026}
\acmYear{2026}
\setcopyright{cc}
\setcctype{by}
\acmConference[CHI EA '26]{Extended Abstracts of the 2026 CHI Conference on Human Factors in Computing Systems}{April 13--17, 2026}{Barcelona, Spain}
\acmBooktitle{Extended Abstracts of the 2026 CHI Conference on Human Factors in Computing Systems (CHI EA '26), April 13--17, 2026, Barcelona, Spain}
\acmDOI{10.1145/3772363.3798959}
\acmISBN{979-8-4007-2281-3/2026/04}

\usepackage{caption}
\usepackage{subcaption}
\usepackage{graphicx}
\usepackage{tabularx}
\usepackage{array}
\usepackage{booktabs}
\begin{document}

\title{SAGE: Sensor-Augmented Grounding Engine for LLM-Powered Sleep Care Agent}

\author{Hansoo Lee}
\authornote{These authors are joint lead authors. Hansoo Lee is also affiliated with KIST, Seoul, Republic of Korea.} 
\affiliation{%
  \institution{Imperial College London}
  \city{London}
  \country{United Kingdom}}
\email{h.lee1@imperial.ac.uk}

\author{Yoonjae Cho}
\authornotemark[1]
\affiliation{%
  \institution{KIST}
  \city{Seoul}
  \country{Republic of Korea}}
\email{yjcho@kist.re.kr}

\author{Sonya S. Kwak}
\authornote{Corresponding author.} 
\affiliation{%
  \institution{KIST}
  \city{Seoul}
  \country{Republic of Korea}}
\email{sonakwak@kist.re.kr}

\author{Rafael A. Calvo}
\authornotemark[2]
\affiliation{%
  \institution{Imperial College London}
  \city{London}
  \country{United Kingdom}}
\email{r.calvo@imperial.ac.uk}

\renewcommand{\shortauthors}{Lee et al.}

\begin{abstract}
Sleep is vital for health, yet access to data alone does not guarantee improvement. While wearables and health apps enable tracking, users face a ``Data-Action Gap,'' struggling to interpret metrics and translate them into action. Current interventions fail to bridge this: static dashboards lack context, rule-based agents rely on rigid scripts, and LLM-agents lack grounding in personal data, causing trust issues. We propose SAGE (Sensor-Augmented Grounding Engine) for an LLM-powered sleep care agent. SAGE normalizes continuous sleep, physiological, and activity data from the sensors into a queryable time-series layer. It supports (1) selective system-initiated monitoring that triggers notifications only upon detecting meaningful deviations against personal baselines to reduce alert fatigue, and (2) user-initiated Q\&A where natural language is translated into executable database queries. By ensuring responses are grounded in precise period, comparison, and metric data, SAGE aims to enhance personalization, traceability, and trust, articulating a novel design space for evidence-based messaging in sleep care.
\end{abstract}



\maketitle

\section{Introduction}
Sleep is essential for physical, cognitive, and mental health~\cite{17ramar2021sleep}. Recently, the proliferation of Consumer Sleep Technology (CST), including smartwatches, mattress sensors, and health apps, has enabled daily sleep tracking outside in everyday, non-clinical settings~\cite{18guillodo2020clinical,21chee2025world}. However, access to data does not guarantee improved health. Users often struggle to interpret variations in accuracy and confidence intervals across devices and metrics~\cite{19lee2023accuracy,20kainec2024evaluating,24manners2025performance,25edouard2021validation}, face a ``Data-Action Gap'' where they fail to translate accumulated data into meaningful action~\cite{22hoang2023knowledge}, or experience anxiety driven by obsessive self-monitoring (orthosomnia)~\cite{23baron2017orthosomnia}. Mainstream technologies, such as static dashboards, fail to support contextual questions like ``How does my sleep compare to last week?''~\cite{22hoang2023knowledge}. Furthermore, they fail to bridge this gap, as they do not provide sufficient confidence intervals or interpretative guidance despite the existence of accuracy variations across devices~\cite{19lee2023accuracy,20kainec2024evaluating,24manners2025performance,25edouard2021validation}.

To fill this gap, various types of conversational agents for sleep support (sleep CAs) have been proposed. Rule-based and scripted sleep CAs lack flexibility, relying on fixed scripts and self-reports~\cite{03werner2019pilot,04philip2020smartphone,05rick2019sleepbot,14subotic2023protocol}, and have focused solely on inducing sleep hygiene behaviors through scheduled notifications, check-in questions, and predefined branching logic~\cite{03werner2019pilot,04philip2020smartphone,05rick2019sleepbot,06su2023assessing}. Even studies on rule/script-based sleep CAs utilizing sensor data used the data merely as intervention triggers rather than as direct evidence for conversation~\cite{06su2023assessing,07tsai2025personalized}. Consequently, user burden increases, and it becomes difficult to explain various sleep metrics in context or provide personalized coaching. Recently, agents utilizing LLMs to support natural conversation and advice for sleep care have been proposed~\cite{08tang2025zzzmate,09alapati2024evaluating, 10bilal2024enhancing, 11mira2024chat}. However, due to a lack of explainable personalization based on the individual's objective status data, they often fail to accurately reflect the period/comparison/metric conditions required by the question. Furthermore, generating answers based on general knowledge can lead to suggestions that do not fit the individual or cause hallucination issues.

To overcome this, recent studies on Sensor-Coupled Personal Health LLMs have attempted to inject wearable data into LLMs by converting it into text (Textualization)~\cite{01wang2025exploring,02khasentino2025personal,13fang2024physiollm}. However, since the meaning of sleep data changes dynamically depending on the time range (e.g., last 7 days vs. last month) and comparison criteria~\cite{22hoang2023knowledge}, this ``Static Summary'' approach incurs high token costs and latency, failing to support ``Dynamic Querying'' that responds to variable user questions~\cite{02khasentino2025personal,15jin2024timellm, 16pmlr-v248-kim24b}. Therefore, effective sleep care requires a ``Continuous Data Grounding'' structure that translates questions into specific queries based on continuous data and provides evidence based explaination.

To address these challenges, we propose \textbf{Sensor-Augmented Grounding Engine (SAGE)}. SAGE normalizes continuous sleep and activity data into a \textbf{``Queryable Time-Series Layer''} rather than converting it into text. Through this, it translates users' natural language questions into executable DB queries to retrieve and explain accurate evidence. This system provides reliable sleep care by supporting not only (1) user-initiated Q\&A but also (2) selective monitoring, which minimizes alert fatigue by intervening only when meaningful deviations against personal baselines are detected.

The contributions of this study are as follows:

\begin{itemize}
    \item \textbf{Proposal of sensor-augmented grounding engine:} We propose a grounding workflow (SAGE) that connects natural language questions to explicit data queries and evidence summaries based on continuous sleep, physiological, and activity data.
    
    \item \textbf{Data-driven Q\&A reflecting period, comparison, and metrics:} By structuring questions into ``When / Compared to What / Which Metric'' to retrieve data and providing explanations based on the results, we enhance the basis for personalization and traceability.
    
    \item \textbf{Design implications for long-term care interaction:} We outline the role division between system-initiated and user-initiated modes and principles for preventing over-intervention, such as intervening only when necessary and treating silence as a deliberate strategy.
\end{itemize}

\section{SAGE Architecture}

\begin{figure}[htbp] 
    \centering
    \includegraphics[width=\linewidth]{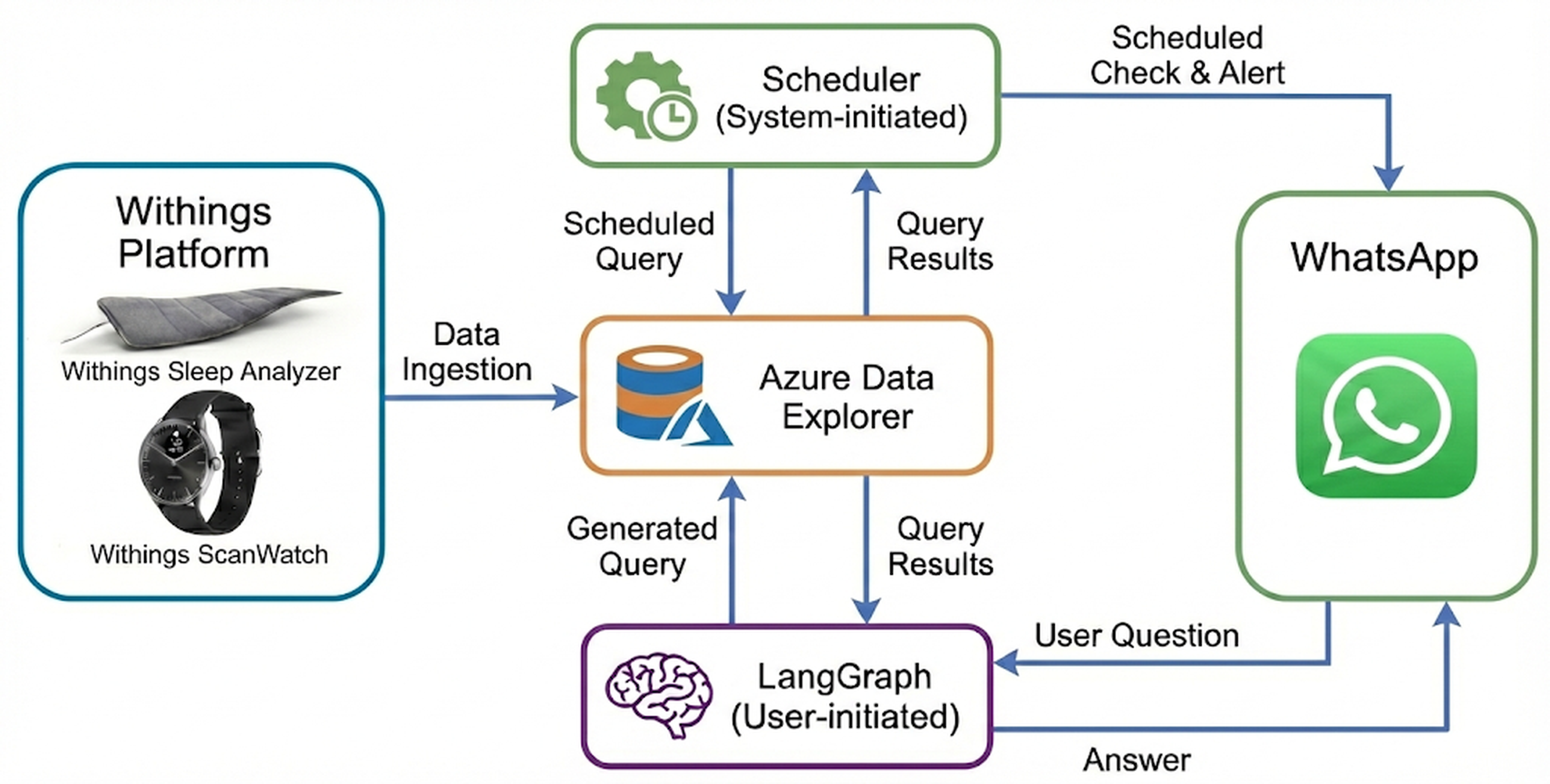} 
    \caption{SAGE Architecture for Sleep Care Agents: Bridges sensors and LLMs via a queryable layer to enable grounded Q\&A and proactive monitoring}
  \Description{Block diagram of the SAGE system architecture showing sensor data ingestion, a queryable time-series layer, and reactive Q&A plus proactive monitoring workflows.}
    \label{fig1:architecture}
\end{figure}

Figure~\ref{fig1:architecture} illustrates the overall interactive architecture of SAGE. The system comprises three main components: \textbf{(1) a data pipeline} that processes real-time sensor data and ingests it into Azure Data Explorer (ADX)~\cite{ref29_microsoft_adx_overview}; \textbf{(2) a conversational interface} that transforms natural language questions into executable data queries using commercial LLMs (e.g., GPT-5.2, GPT-5 mini~\cite{ref37_openai_latest_model_guide}) and LangGraph~\cite{ref33_langchain_langgraph_overview}; and \textbf{(3) a monitoring module} that provides selective notifications when significant deviations are detected.

\subsection{Sensor Ingestion \& Queryable Time-Series Layer}

In this layer, data from the Withings Sleep Analyzer~\cite{ref32_withings_sleep_analyzer_support_section} and ScanWatch~\cite{ref36_withings_scanwatch_support_section} are ingested into ADX and normalized into ‘queryable time-series data.’ We adopted ADX over relational databases (RDBs) because it is optimized for the real-time ingestion and analysis of large-scale time-series data, offering superior performance in complex time-based aggregations and comparison operations~\cite{ref30_microsoft_adx_ingest_overview,ref31_microsoft_kusto_update_policy}. Specifically, prioritizing the \textbf{continuity} and \textbf{depth} of data collection over the wearable devices used in prior studies~\cite{01wang2025exploring,12_10.1093/jamiaopen/ooaf067,13fang2024physiollm, ref39_samsung_gear_sport_product}(e.g., Fitbit~\cite{ref34_fitbit_sense_product}, Oura Ring~\cite{ref35_oura_help_center}), we utilized the Withings ecosystem. The Sleep Analyzer, installed under the mattress as a `nearable' device, is \textbf{zero-burden} (requiring no charging or wearing), preventing data loss due to non-wear—a chronic issue with wearables—while providing clinical-grade deep metrics such as apnea and snoring~\cite{24manners2025performance,25edouard2021validation, 27article,ref40_withings_sleep_analyzer_user_guide_v6}. Additionally, the ScanWatch, with its 30-day battery life, continuously collects daytime activity data (e.g., steps, heart rate), enabling an integrated analysis of the \textbf{activity-sleep context} (i.e., how daytime activity affects sleep efficiency)~\cite{ref28_withings_scanwatch_battery}.

Raw data received via the Withings Cloud API in JSON format is automatically converted into structured tables via \textbf{ADX update policies}. Activity data is transformed into daily numeric fields (steps, calories, avg/min/max heart rate), while sleep data is normalized into session-based tables containing sleep stages (Light/Deep/REM), Wake After Sleep Onset (WASO), and respiratory metrics. During this process, all timestamps are normalized to the user’s time zone, and the fields are cast to computable types. Importantly, this normalization defines the storage schema but does not restrict query resolution: depending on the user’s question, SAGE dynamically generates queries that retrieve raw records or compute aggregates over the appropriate timeframe. This structured pre-processing ensures \textbf{traceability} by allowing the LLM to retrieve precise DB values rather than inferring numbers arbitrarily.

\subsection{Grounded Query Generation: From Questions to Queries}

\begin{figure*}[htbp]
    \centering
    \includegraphics[width=0.8\linewidth]{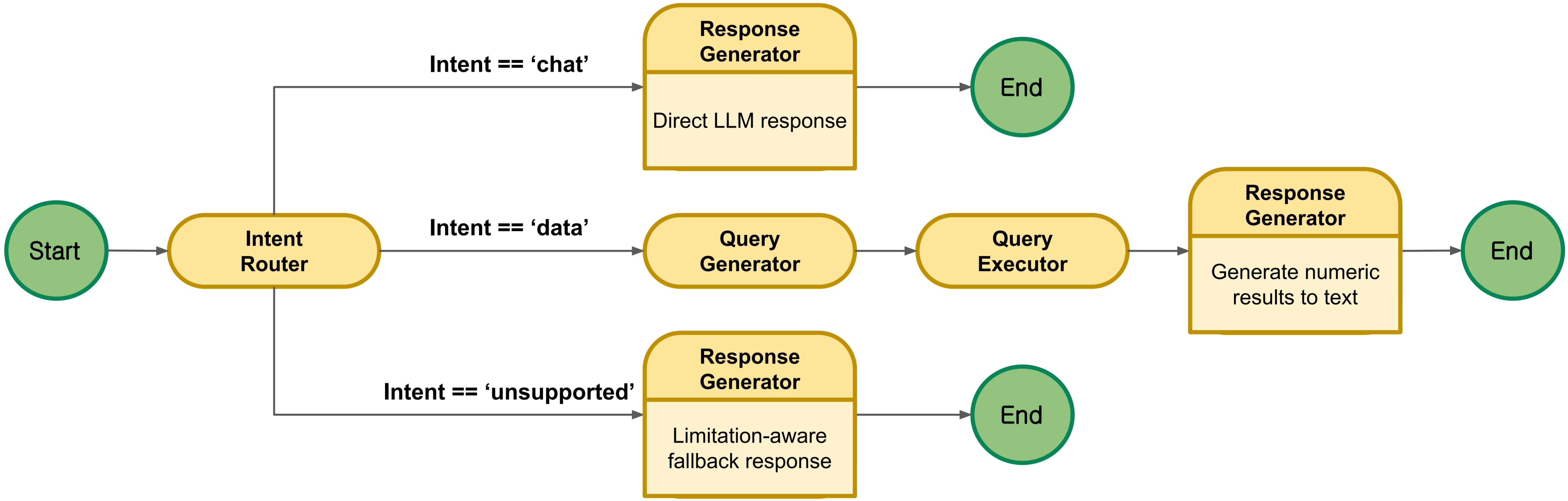} 
    \caption{Intent Routing Logic: A LangGraph state machine that classifies inputs into chat or data paths to prevent hallucinations}
  \Description{State-machine diagram for intent routing: a user message is classified into chat, data, or unsupported; the data branch generates a database query, retrieves results, and produces a grounded response.}
    \label{fig2:langgraph}
\end{figure*}

\begin{figure*}[htbp]
  \centering

  \begin{subfigure}[t]{0.32\linewidth}
    \centering
    \includegraphics[width=\linewidth,height=2.3cm]{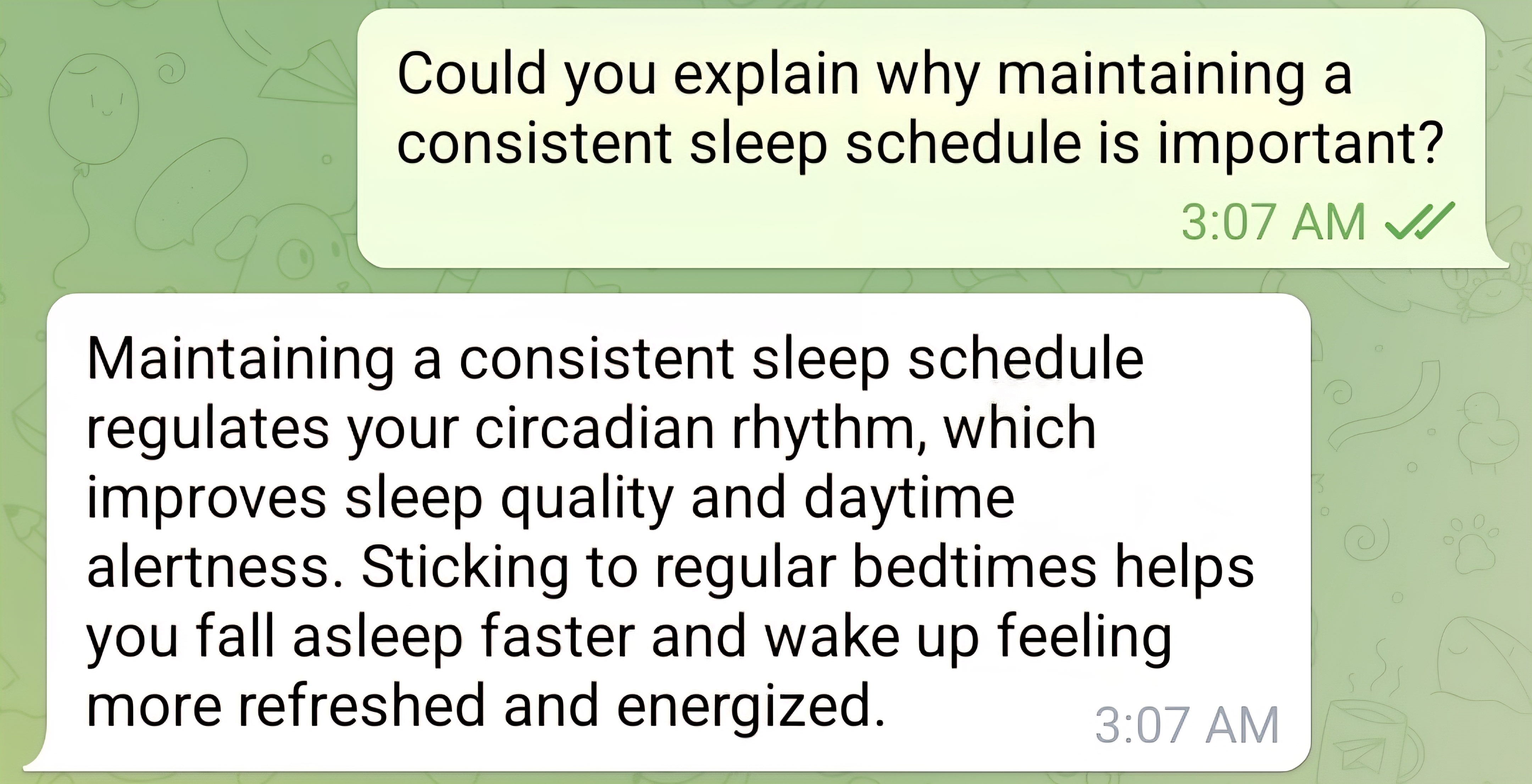}
    \caption{General conversation}
    \label{fig3a:msg-a}
  \end{subfigure}\hfill
  \begin{subfigure}[t]{0.32\linewidth}
    \centering
    \includegraphics[width=\linewidth,height=2.3cm]{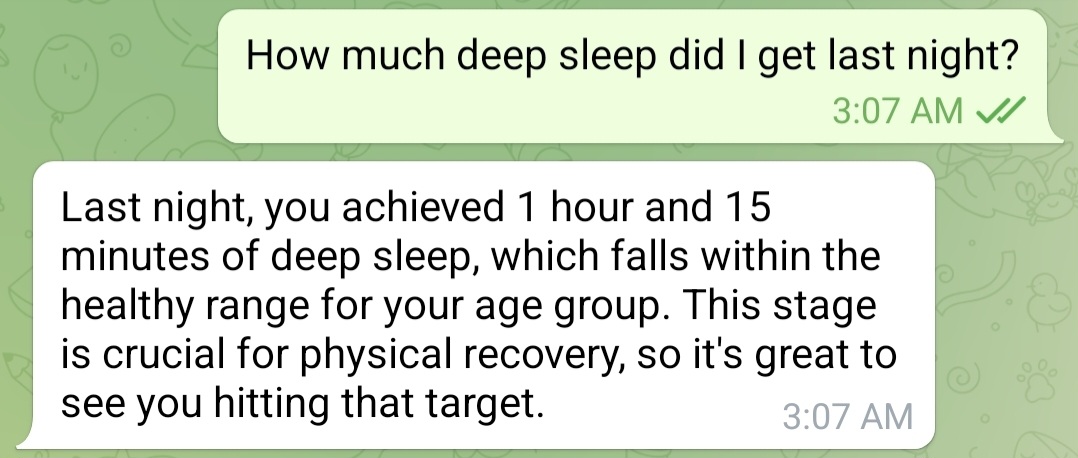}
    \caption{Simple data retrieval}
    \label{fig3b:msg-b}
  \end{subfigure}\hfill
  \begin{subfigure}[t]{0.32\linewidth}
    \centering
    \includegraphics[width=\linewidth,height=2.3cm]{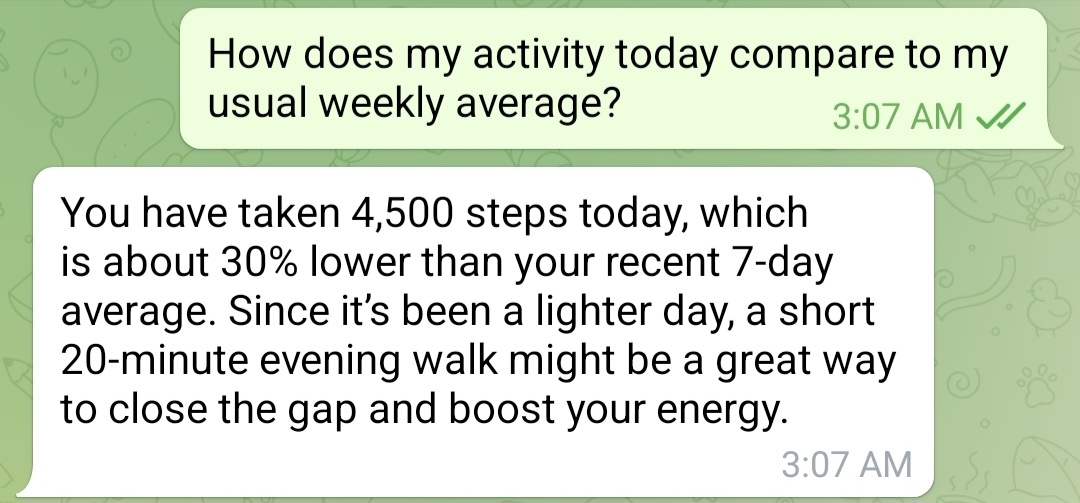}
    \caption{Comparative data query}
    \label{fig3c:msg-c}
  \end{subfigure}

  \par\smallskip

  \begin{subfigure}[t]{0.32\linewidth}
    \centering
    \includegraphics[width=\linewidth,height=2.3cm]{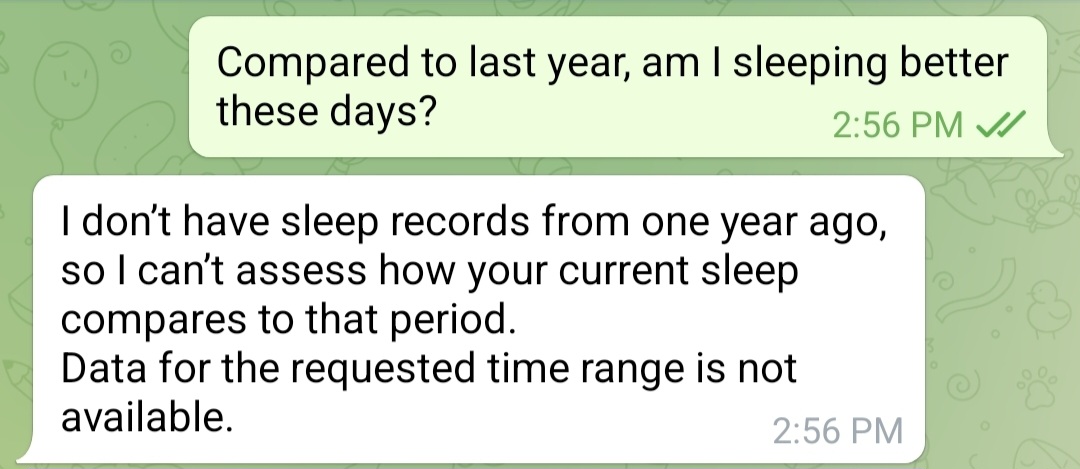}
    \caption{Null-data request}
    \label{fig3d:msg-d}
  \end{subfigure}\hfill
  \begin{subfigure}[t]{0.32\linewidth}
    \centering
    \includegraphics[width=\linewidth,height=2.3cm]{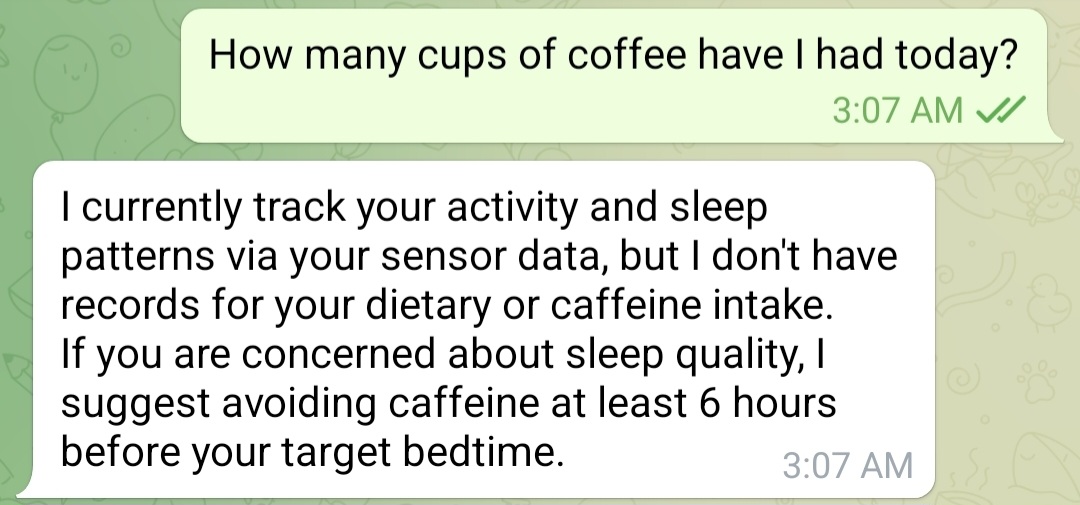}
    \caption{Unsupported/ambiguous request}
    \label{fig3e:msg-e}
  \end{subfigure}\hfill
  \begin{subfigure}[t]{0.32\linewidth}
    \centering
    \includegraphics[width=\linewidth,height=2.3cm]{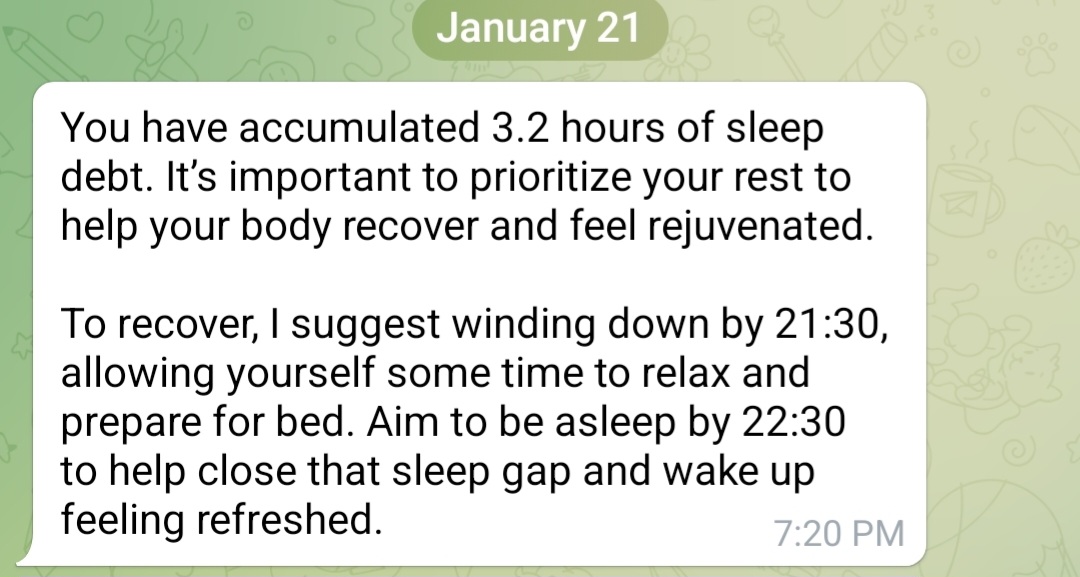}
    \caption{System-initiated notification}
    \label{fig3f:msg-f}
  \end{subfigure}

  \caption{Example of messaging interactions of the SAGE-based conversational sleep care agent}
  \Description{Six messaging screenshots arranged in a 2-by-3 grid: (a) general conversation, (b) simple data retrieval, (c) comparative query, (d) null-data request, (e) unsupported or ambiguous request, and (f) system-initiated notification.}
  \label{fig3:messaging-examples}
\end{figure*}

The core of the user-initiated interaction is converting natural language questions into actionable data requests and providing grounded explanations. Controlled by a LangGraph-based state graph (Figure~\ref{fig2:langgraph}), user inputs are classified by an \textbf{intent router} into three categories: chat, data, or unsupported. The system interface employs WhatsApp~\cite{42_whatsapp_business_platform}, integrated via Twilio~\cite{41_twilio_whatsapp_api} and Azure Functions~\cite{43_microsoft_functions_overview} for real-time message relay.

The first case involves intents classified as \textit{\textbf{`chat'}}. As illustrated in Figure~\ref{fig3a:msg-a}—where the user asks, "Could you explain why maintaining a consistent sleep schedule is important?"—this category includes casual conversations or requests for general medical advice that do not require specific data retrieval. In this path, the system utilizes the \textbf{response generator} to produce a \textbf{direct LLM response} based on pre-trained knowledge.

The second case involves intents classified as \textit{\textbf{`data'}}, covering queries that require verification of actual sensor data, as shown in Figure~\ref{fig3b:msg-b}--\ref{fig3d:msg-d}. Upon entering this path, the \textbf{query generator} converts natural language inquiries into Kusto Query Language (KQL) executable on ADX. A primary reason for adopting ADX over RDBs is that KQL’s pipeline-based (|) sequential logic is structurally better suited for the LLM’s step-by-step reasoning compared to SQL’s frequent nested queries, thereby minimizing syntax errors. Subsequently, the \textbf{response generator} adapts its output strategy based on the specific query type. For \textbf{simple data retrieval}, such as ``How much deep sleep did I get last night?'' in Figure~\ref{fig3b:msg-b}, the system directly returns the retrieved value (1 hour and 15 minutes). Similarly, for \textbf{comparative data queries} like "How does my activity today compare to my usual weekly average?" in Figure 3(c), it provides an analysis contrasting current data (4,500 steps) with historical averages. Conversely, for \textbf{null-data requests} where data is unavailable for the queried period, as depicted in Figure 3(d), the system explicitly states that ``no records exist'' to prevent hallucinations.

The third case is when the intent is classified as \textbf{`unsupported'}. This applies to out-of-scope queries, such as the user asking about coffee intake in Figure~\ref{fig3e:msg-e}, which the system does not track. In such cases, the response generator creates a response informing the user that the question cannot be answered within the current data collection scope.

Unlike previous sensor-coupled LLM studies~\cite{01wang2025exploring,02khasentino2025personal,13fang2024physiollm} that textualize time-series data for model injection, we propose an \textbf{``LLM → DB query generation''} approach. Direct injection of long time-series data incurs high token costs and latency. More importantly, allowing a probabilistic LLM to perform arithmetic carries a high risk of hallucination, potentially presenting non-existent patterns as facts—a severe ethical issue in healthcare~\cite{26lee2023benefits}. Therefore, we establish ADX as the \textbf{single source of truth}, confining the LLM to the roles of `question interpretation' and `explanation of retrieved facts' to ensure both computational accuracy and ethical safety.

\subsection{Proactive \& Selective Monitoring}

SAGE provides a \textbf{system-initiated notification} mechanism beyond passive responses. A daily scheduler executes \textbf{predefined query templates} to monitor data, avoiding the cost and instability of generating queries on the fly. Notifications are triggered selectively only when significant deviations (e.g., >±30\% change from the 7-day moving average) or critical sleep debts are detected, as shown in Figure~\ref{fig3f:msg-f}. This design aims to minimize \textbf{notification fatigue} by avoiding mechanical repetitive reports and intervening only when meaningful changes in user status occur.

\section{Preliminary Technical Validation}

We conducted an initial technical validation to examine whether the proposed SAGE system can reliably execute its designed pipeline (\textbf{query generation → data retrieval → response generation}). This evaluation focuses on verifying the structural and functional correctness of the end-to-end process.

\subsection{Test Dataset Construction}
The test dataset reflected query types observed in real-world messaging environments, comprising three categories in Figure~\ref{fig3b:msg-b}--\ref{fig3d:msg-d}: \textbf{(1) simple data retrieval}, \textbf{(2) comparative query}, and \textbf{(3) null-data request}. We conducted the test with 30 questions for each category, synthesized via an LLM seeded with representative examples. This distribution was designed to observe which stage of pipeline, intent parsing, query generation, or data interpretation, encounters difficulties relative to query complexity. Conversely, \textbf{general conversation} (Figure~\ref{fig3a:msg-a}), \textbf{unsupported data request} (Figure~\ref{fig3e:msg-e}), and \textbf{system-initiated notification} (Figure~\ref{fig3f:msg-f}) were excluded from the evaluation. These interaction types were omitted because they tend to operate stably by either not traversing the end-to-end data retrieval path or being processed via predefined logic.

\subsection{Pipeline Validation Metrics (M1--M4)}

The evaluation employed the following four metrics to verify the entire pipeline, ranging from KQL executability to the accuracy of the final response.
\begin{itemize}
    \item \textbf{M1 (Executable query rate):} This measures whether the LLM-generated KQL executes in the ADX environment without syntax errors, thereby assessing the stability of the query generation module.
    
    \item \textbf{M2 (Query intent match):} Regardless of M1's execution success, this evaluates whether the generated KQL structure logically and correctly reflects the user's core query intent (e.g., target period, comparison criteria).
    
    \item \textbf{M3 (Data retrieval correctness):} Applied to queries that passed M1, this verifies the accuracy of numerical calculation by checking whether the results returned by ADX match the manually calculated ground truth.

    \item \textbf{M4 (Response faithfulness):} Also applied to queries that passed M1, this confirms whether the final textual response describes the retrieved data without distortion.
    
\end{itemize}
In particular, this evaluation adopts a conditional approach where, even if an intent mismatch occurs in M2, M3, and M4 are considered passed if the executed KQL yields the correct result. This staged design allows for the clear \textbf{decoupling} and identification of failure causes, determining whether they stem from syntax (M1), intent (M2), data retrieval (M3), or response generation (M4).

\subsection{Results and Failure Analysis}

Experimental results demonstrated that the proposed pipeline operated stably with 100\% success in M1–M4 for simple retrieval and most comparative queries. However, errors occurred in complex period comparisons and multi-condition queries due to structural limitations, identifying three main failure modes.
\begin{itemize}
    \item First, in queries requiring simultaneous comparison of two distinct periods (e.g., ``this week vs. last week''), \textbf{syntax errors (M1 error)} occurred because the query operation order was incorrectly constructed to perform aggregation before time filtering. However, we plan to address this by explicitly enforcing a \textbf{``filter-then-aggregate'' rule} at the prompt level. 
    
    \item Second, when relative time expressions involving the present (e.g., ``this week'') were used, \textbf{semantic errors (M2 error)} were observed where the generated query range incorrectly extended into the future. This is attributed to the instability of natural language time interpretation and can be mitigated by implementing a \textbf{time normalization module}. 
    
    \item Lastly, when values for the same metric differed between two devices for the same period, \textbf{logic errors (M2 error)} occurred where duplicate data were aggregated. This stems from unclear data source selection and integration rules; we intend to resolve this via \textbf{sensor priority policies or an integrated metric schema}.
\end{itemize}

\section{Discussion}

\subsection{HCI Implications for Sleep Care Agents}
SAGE contributes to HCI by enabling evidence-based sleep care that resolves the trade-off between the rigidity of rule-based systems and the lack of grounding in generic LLMs. While rule-based systems fail to handle flexible comparison queries (e.g., ``vs last month'') and generic LLMs risk hallucinations, SAGE converts natural language into explicit data requests (period, comparison, and metrics) to provide actionable explanations at ``moments of curiosity.'' While standard sleep tracking apps effectively use visual dashboards (e.g., hypnograms) to show broad trends, users often struggle to interpret these graphs into actionable steps, leading to a ``Data-Action Gap.'' SAGE does not aim to replace these visuals, but rather to complement them by using a conversational messaging interface (e.g., WhatsApp) to deliver precise, context-aware textual explanations and actionable nudges that static graphs cannot easily convey. Crucially, SAGE minimizes \textbf{alert fatigue} through \textbf{selective monitoring,} intervening only when meaningful deviations occur, rather than overwhelming users with constant notifications. 

Furthermore, unlike prior sensor-coupled LLMs that rely on static text summaries or end-to-end sensor fusion models, SAGE adopts \textbf{DB-centered grounding}. This approach aim to ensure \textbf{auditability} by externalizing the numerical basis of answers and improves \textbf{scalability}, allowing new sensors/metrics without model retraining, which we anticipate as key factors for trust and burden minimization in sleep care, to be verified in future user studies. More broadly, this architectural choice suggests that grounding can be treated as a structural property of conversational health systems rather than solely as a safeguard against hallucination, offering implications for future evidence-based agent design beyond the sleep domain. 

\subsection{Limitations \& Future Work}
Currently, SAGE has limited mechanism for multi-turn clarification of ambiguous queries and lacks integration of relevant contextual factors, such as caffeine intake and stress. Future work will focus on: (a) designing efficient multi-turn interactions to resolve ambiguity with minimal friction, (b) integrating multi-modal contexts (activity, diet, schedules) to explore quasi-causal ``behavior-sleep'' relationships safely, and (c) refining personalized deviation detection to prevent over-intervention. Technically, we plan to improve query generation robustness against edge cases. Furthermore, while SAGE transitions from static text summaries to a dynamic queryable layer, it still relies on API-level aggregations (e.g., session-based sleep stages, daily heart rate extremes) rather than continuous, beat-to-beat raw physiological signals. We acknowledge this loss of detailed signal information as a necessary trade-off to ensure computational efficiency and lower latency in a real-time conversational agent.

Additionally, as our current technical validation relied on synthetic queries, it limits our understanding of real-world user behavior and perceived trustworthiness. Furthermore, the current text-only messaging interface may impose cognitive load when conveying complex sleep trends. Therefore, we plan to integrate lightweight visual aids (e.g., hypnograms and trend charts) alongside text explanations to improve intuitiveness. Finally, long-term field studies with real users are required to evaluate explainability, personalization impacts, and perceived trust. To comprehensively assess SAGE’s usability and clinical efficacy, we will compare our system against three baselines: (1) traditional rule-based agents, (2) general knowledge-based LLM agents, and (3) standard wearable sensor-based health dashboards that rely primarily on visualizations (e.g., Apple Health, Samsung Health, Google Fit).

\section{Conclusion}
We presented an initial model of a sensor-grounded sleep care agent that transforms user inquiries into precise data requests and evidence-based explanations based on continuous sleep and activity data. The decoupling of a \textbf{``queryable time-series layer''} from the conversational agent offers a practical framework for transforming passive sleep tracking into interactive, messaging-based care.

\begin{acks}
This project was partially supported by the Sejong Science Fellowship [RS-2025-00559234], funded by the National Research Foundation (NRF); the Design Innovation Program [RS-2025-14343002], funded by the Ministry of Trade, Industry and Energy (MOTIE); the Leverhulme Trust through the Leverhulme Centre for the Future of Intelligence [RC-2015-067]; and the UK Dementia Research Institute (award number UK DRI-7005) through UK DRI Ltd, which is principally funded by the Medical Research Council and additional funding partner, the Alzheimer’s Society.
\end{acks}

\bibliographystyle{ACM-Reference-Format}
\bibliography{sample-base}

@inproceedings{01wang2025exploring,
  title={Exploring Personalized Health Support through Data-Driven, Theory-Guided LLMs: A Case Study in Sleep Health},
  author={Wang, Xingbo and Griffith, Janessa and Adler, Daniel A and Castillo, Joey and Choudhury, Tanzeem and Wang, Fei},
  booktitle={Proceedings of the 2025 CHI Conference on Human Factors in Computing Systems},
  pages={1--15},
  year={2025}
}

@article{02khasentino2025personal,
  title={A personal health large language model for sleep and fitness coaching},
  author={Khasentino, Justin and Belyaeva, Anastasiya and Liu, Xin and Yang, Zhun and Furlotte, Nicholas A and Lee, Chace and Schenck, Erik and Patel, Yojan and Cui, Jian and Schneider, Logan Douglas and others},
  journal={Nature Medicine},
  volume={31},
  number={10},
  pages={3394--3403},
  year={2025},
  publisher={Nature Publishing Group US New York}
}

@article{03werner2019pilot,
  title={Pilot evaluation of the Sleep Ninja: a smartphone application for adolescent insomnia symptoms},
  author={Werner-Seidler, Aliza and Wong, Quincy and Johnston, Lara and O’Dea, Bridianne and Torok, Michelle and Christensen, Helen},
  journal={BMJ open},
  volume={9},
  number={5},
  pages={e026502},
  year={2019},
  publisher={British Medical Journal Publishing Group}
}

@article{04philip2020smartphone,
  title={Smartphone-based virtual agents to help individuals with sleep concerns during COVID-19 confinement: feasibility study},
  author={Philip, Pierre and Dupuy, Lucile and Morin, Charles M and de Sevin, Etienne and Bioulac, St{\'e}phanie and Taillard, Jacques and Serre, Fuschia and Auriacombe, Marc and Micoulaud-Franchi, Jean-Arthur},
  journal={Journal of medical Internet research},
  volume={22},
  number={12},
  pages={e24268},
  year={2020},
  publisher={JMIR Publications Toronto, Canada}
}

@inproceedings{05rick2019sleepbot,
  title={SleepBot: encouraging sleep hygiene using an intelligent chatbot},
  author={Rick, Steven R and Goldberg, Aaron Paul and Weibel, Nadir},
  booktitle={Companion Proceedings of the 24th International Conference on Intelligent User Interfaces},
  pages={107--108},
  year={2019}
}

@article{06su2023assessing,
  title={Assessing a Sleep Interviewing chatbot to improve subjective and objective sleep: Protocol for an Observational Feasibility Study},
  author={Su, Ting and Calvo, Rafael A and Jouaiti, Melanie and Daniels, Sarah and Kirby, Pippa and Dijk, Derk-Jan and Della Monica, Ciro and Vaidyanathan, Ravi},
  journal={JMIR research protocols},
  volume={12},
  number={1},
  pages={e45752},
  year={2023},
  publisher={JMIR Publications Inc., Toronto, Canada}
}

@article{07tsai2025personalized,
  title={A Personalized, Texting-Based Conversational Agent to Address Sleep Disturbance in Individuals Who Have Survived Breast Cancer: Protocol for a Pilot Waitlist Randomized Controlled Trial},
  author={Tsai, Chi-shan and Szewczyk, Warren and Drerup, Michelle and Liao, Jason and Vasbinder, Alexi and Greenlee, Heather and Heffner, Jaimee L and Yung, Rachel and Reding, Kerryn W},
  journal={JMIR Research Protocols},
  volume={14},
  number={1},
  pages={e62712},
  year={2025},
  publisher={JMIR Publications Inc., Toronto, Canada}
}

@inproceedings{08tang2025zzzmate,
  title={ZzzMate: A Self-Conscious Emotion-Aware Chatbot for Sleep Intervention},
  author={Tang, Xiao and Li, Zhuying and Sun, Xin and Xu, Xuhai and Zhang, Min-Ling},
  booktitle={Proceedings of the Extended Abstracts of the CHI Conference on Human Factors in Computing Systems},
  pages={1--7},
  year={2025}
}

@article{09alapati2024evaluating,
  title={Evaluating insomnia queries from an artificial intelligence chatbot for patient education},
  author={Alapati, Rahul and Campbell, Daniel and Molin, Nicole and Creighton, Erin and Wei, Zhikui and Boon, Maurits and Huntley, Colin},
  journal={Journal of Clinical Sleep Medicine},
  volume={20},
  number={4},
  pages={583--594},
  year={2024},
  publisher={American Academy of Sleep Medicine}
}

@article{10bilal2024enhancing,
  title={Enhancing awareness and self-diagnosis of obstructive sleep apnea using AI-powered Chatbots: the role of ChatGPT in revolutionizing healthcare},
  author={Bilal, Maham and Jamil, Yumna and Rana, Dua and Shah, Hussain Haider},
  journal={Annals of biomedical engineering},
  volume={52},
  number={2},
  pages={136--138},
  year={2024},
  publisher={Springer}
}

@article{11mira2024chat,
  title={Chat GPT for the management of obstructive sleep apnea: do we have a polar star?},
  author={Mira, Felipe Ahumada and Favier, Valentin and dos Santos Sobreira Nunes, Heloisa and de Castro, Joana Vaz and Carsuzaa, Florent and Meccariello, Giuseppe and Vicini, Claudio and De Vito, Andrea and Lechien, Jerome R and Chiesa-Estomba, Carlos and others},
  journal={European Archives of Oto-Rhino-Laryngology},
  volume={281},
  number={4},
  pages={2087--2093},
  year={2024},
  publisher={Springer}
}

@article{12_10.1093/jamiaopen/ooaf067,
  title={Conversational health agents: a personalized large language model-powered agent framework},
  author={Abbasian, Mahyar and Azimi, Iman and Rahmani, Amir M and Jain, Ramesh},
  journal={JAMIA Open},
  volume={8},
  number={4},
  pages={ooaf067},
  year={2025},
  publisher={Oxford University Press}
}

@inproceedings{13fang2024physiollm,
  title={Physiollm: Supporting personalized health insights with wearables and large language models},
  author={Fang, Cathy Mengying and Danry, Valdemar and Whitmore, Nathan and Bao, Andria and Hutchison, Andrew and Pierce, Cayden and Maes, Pattie},
  booktitle={2024 IEEE EMBS International Conference on Biomedical and Health Informatics (BHI)},
  pages={1--8},
  year={2024},
  organization={IEEE}
}

@article{14subotic2023protocol,
  title={Protocol for a randomised controlled trial evaluating the effect of a CBT-I smartphone application (Sleep Ninja{\textregistered}) on insomnia symptoms in children},
  author={Subotic-Kerry, M and Werner-Seidler, A and Corkish, B and Batterham, PJ and Sicouri, G and Hudson, J and Christensen, H and O’dea, B and Li, SH},
  journal={BMC psychiatry},
  volume={23},
  number={1},
  pages={684},
  year={2023},
  publisher={Springer}
}

@inproceedings{
15jin2024timellm,
title={Time-{LLM}: Time Series Forecasting by Reprogramming Large Language Models},
author={Ming Jin and Shiyu Wang and Lintao Ma and Zhixuan Chu and James Y. Zhang and Xiaoming Shi and Pin-Yu Chen and Yuxuan Liang and Yuan-Fang Li and Shirui Pan and Qingsong Wen},
booktitle={The Twelfth International Conference on Learning Representations},
year={2024},
url={https://openreview.net/forum?id=Unb5CVPtae}
}

@InProceedings{16pmlr-v248-kim24b,
  title = 	 {Health-LLM: Large Language Models for Health Prediction via Wearable Sensor Data},
  author =       {Kim, Yubin and Xu, Xuhai and McDuff, Daniel and Breazeal, Cynthia and Park, Hae Won},
  booktitle = 	 {Proceedings of the fifth Conference on Health, Inference, and Learning},
  pages = 	 {522--539},
  year = 	 {2024},
  editor = 	 {Pollard, Tom and Choi, Edward and Singhal, Pankhuri and Hughes, Michael and Sizikova, Elena and Mortazavi, Bobak and Chen, Irene and Wang, Fei and Sarker, Tasmie and McDermott, Matthew and Ghassemi, Marzyeh},
  volume = 	 {248},
  series = 	 {Proceedings of Machine Learning Research},
  month = 	 {27--28 Jun},
  publisher =    {PMLR},
  url = 	 {https://proceedings.mlr.press/v248/kim24b.html}
}

@article{17ramar2021sleep,
  title={Sleep is essential to health: an American Academy of Sleep Medicine position statement},
  author={Ramar, Kannan and Malhotra, Raman K and Carden, Kelly A and Martin, Jennifer L and Abbasi-Feinberg, Fariha and Aurora, R Nisha and Kapur, Vishesh K and Olson, Eric J and Rosen, Carol L and Rowley, James A and others},
  journal={Journal of Clinical Sleep Medicine},
  volume={17},
  number={10},
  pages={2115--2119},
  year={2021},
  publisher={American Academy of Sleep Medicine}
}

@article{18guillodo2020clinical,
  title={Clinical applications of mobile health wearable--based sleep monitoring: systematic review},
  author={Guillodo, Elise and Lemey, Christophe and Simonnet, Mathieu and Walter, Michel and Baca-Garc{\'\i}a, Enrique and Masetti, Vincent and Moga, Sorin and Larsen, Mark and HUGOPSY Network and Ropars, Juliette and others},
  journal={JMIR mHealth and uHealth},
  volume={8},
  number={4},
  pages={e10733},
  year={2020},
  publisher={JMIR Publications Toronto, Canada}
}

@article{19lee2023accuracy,
  title={Accuracy of 11 wearable, nearable, and airable consumer sleep trackers: prospective multicenter validation study},
  author={Lee, Taeyoung and Cho, Younghoon and Cha, Kwang Su and Jung, Jinhwan and Cho, Jungim and Kim, Hyunggug and Kim, Daewoo and Hong, Joonki and Lee, Dongheon and Keum, Moonsik and others},
  journal={JMIR mHealth and uHealth},
  volume={11},
  number={1},
  pages={e50983},
  year={2023},
  publisher={JMIR Publications Inc., Toronto, Canada}
}

@article{20kainec2024evaluating,
  title={Evaluating accuracy in five commercial sleep-tracking devices compared to research-grade actigraphy and polysomnography},
  author={Kainec, Kyle A and Caccavaro, Jamie and Barnes, Morgan and Hoff, Chloe and Berlin, Annika and Spencer, Rebecca MC},
  journal={Sensors},
  volume={24},
  number={2},
  pages={635},
  year={2024},
  publisher={MDPI}
}

@article{21chee2025world,
  title={World Sleep Society recommendations for the use of wearable consumer health trackers that monitor sleep},
  author={Chee, Michael WL and Baumert, Mathias and Scott, Hannah and Cellini, Nicola and Goldstein, Cathy and Baron, Kelly and Imtiaz, Syed A and Penzel, Thomas and Kushida, Clete A and others},
  journal={Sleep Medicine},
  pages={106506},
  year={2025},
  publisher={Elsevier}
}

@article{22hoang2023knowledge,
  title={Knowledge discovery in ubiquitous and personal sleep tracking: scoping review},
  author={Hoang, Nhung Huyen and Liang, Zilu},
  journal={JMIR mHealth and uHealth},
  volume={11},
  pages={e42750},
  year={2023},
  publisher={JMIR Publications Toronto, Canada}
}

@article{23baron2017orthosomnia,
  title={Orthosomnia: are some patients taking the quantified self too far?},
  author={Baron, Kelly Glazer and Abbott, Sabra and Jao, Nancy and Manalo, Natalie and Mullen, Rebecca},
  journal={Journal of clinical sleep medicine},
  volume={13},
  number={2},
  pages={351--354},
  year={2017},
  publisher={American Academy of Sleep Medicine}
}

@article{24manners2025performance,
  title={Performance evaluation of an under-mattress sleep sensor versus polysomnography in> 400 nights with healthy and unhealthy sleep},
  author={Manners, Jack and Kemps, Eva and Lechat, Bastien and Catcheside, Peter and Eckert, Danny J and Scott, Hannah},
  journal={Journal of sleep research},
  volume={34},
  number={6},
  pages={e14480},
  year={2025},
  publisher={Wiley Online Library}
}

@article{25edouard2021validation,
  title={Validation of the Withings Sleep Analyzer, an under-the-mattress device for the detection of moderate-severe sleep apnea syndrome},
  author={Edouard, Paul and Campo, David and Bartet, Pierre and Yang, Rui-Yi and Bruyneel, Marie and Roisman, Gabriel and Escourrou, Pierre},
  journal={Journal of Clinical Sleep Medicine},
  volume={17},
  number={6},
  pages={1217--1227},
  year={2021},
  publisher={American Academy of Sleep Medicine}
}

@article{26lee2023benefits,
  title={Benefits, limits, and risks of GPT-4 as an AI chatbot for medicine},
  author={Lee, Peter and Bubeck, Sebastien and Petro, Joseph},
  journal={New England Journal of Medicine},
  volume={388},
  number={13},
  pages={1233--1239},
  year={2023},
  publisher={Mass Medical Soc}
}

@article{27article,
author = {Donckt, Jonas and Vandenbussche, Nicolas and Donckt, Jeroen and Chen, Stephanie and Stojchevska, Marija and De Brouwer, Mathias and Steenwinckel, Bram and Paemeleire, Koen and Ongenae, Femke and Hoecke, Sofie},
year = {2024},
month = {07},
pages = {},
title = {Mitigating data quality challenges in ambulatory wrist-worn wearable monitoring through analytical and practical approaches},
volume = {14},
journal = {Scientific Reports},
doi = {10.1038/s41598-024-67767-3}
}

@misc{ref28_withings_scanwatch_battery,
  author       = {{Withings}},
  title        = {{ScanWatch}: How long can the battery of my watch last?},
  howpublished = {Withings Support},
  year         = {2026},
  month        = jan,
  url          = {https://support.withings.com/hc/en-us/articles/360009967878-ScanWatch-How-long-can-the-battery-of-my-watch-last},
  note         = {Retrieved January 22, 2026, from the listed URL}
}

@misc{ref29_microsoft_adx_overview,
  author       = {{Microsoft}},
  title        = {What is {Azure Data Explorer}?},
  howpublished = {Microsoft Learn},
  year         = {2025},
  month        = jun,
  url          = {https://learn.microsoft.com/en-us/azure/data-explorer/data-explorer-overview},
  note         = {Page date: June 10, 2025. Retrieved January 22, 2026, from the listed URL}
}

@misc{ref30_microsoft_adx_ingest_overview,
  author       = {{Microsoft}},
  title        = {{Azure Data Explorer} data ingestion overview},
  howpublished = {Microsoft Learn},
  year         = {2025},
  month        = sep,
  url          = {https://learn.microsoft.com/en-us/azure/data-explorer/ingest-data-overview},
  note         = {Page date: September 10, 2025. Retrieved January 22, 2026, from the listed URL}
}

@misc{ref31_microsoft_kusto_update_policy,
  author       = {{Microsoft}},
  title        = {Update policy overview},
  howpublished = {Microsoft Learn},
  year         = {2026},
  month        = jan,
  url          = {https://learn.microsoft.com/en-us/kusto/management/update-policy?view=microsoft-fabric},
  note         = {Page date: January 21, 2026. Retrieved January 22, 2026, from the listed URL}
}

@misc{ref32_withings_sleep_analyzer_support_section,
  author       = {{Withings}},
  title        = {{Sleep Analyzer} ({EU} \& {ROW})},
  howpublished = {Withings Support},
  year         = {2026},
  month        = jan,
  url          = {https://support.withings.com/hc/en-us/sections/6215955439505-Sleep-Analyzer-EU-ROW},
  note         = {Retrieved January 22, 2026, from the listed URL}
}

@misc{ref33_langchain_langgraph_overview,
  author       = {{LangChain}},
  title        = {{LangGraph} overview},
  howpublished = {LangChain Docs},
  year         = {2026},
  month        = jan,
  url          = {https://docs.langchain.com/oss/python/langgraph/overview},
  note         = {Retrieved January 22, 2026, from the listed URL}
}

@misc{ref34_fitbit_sense_product,
  author       = {{Fitbit}},
  title        = {{Fitbit Sense}},
  howpublished = {Fitbit},
  year         = {2026},
  month        = jan,
  url          = {https://www.fitbit.com/sg/sense},
  note         = {Retrieved January 22, 2026, from the listed URL}
}

@misc{ref35_oura_help_center,
  author       = {{Oura}},
  title        = {Oura help center},
  howpublished = {Oura Support},
  year         = {2026},
  month        = jan,
  url          = {https://support.ouraring.com/hc/en-us},
  note         = {Retrieved January 22, 2026, from the listed URL}
}

@misc{ref36_withings_scanwatch_support_section,
  author       = {{Withings}},
  title        = {{ScanWatch}},
  howpublished = {Withings Support},
  year         = {2026},
  month        = jan,
  url          = {https://support.withings.com/hc/en-us/sections/4411036796433-ScanWatch},
  note         = {Retrieved January 22, 2026, from the listed URL}
}

@misc{ref37_openai_latest_model_guide,
  author       = {{OpenAI}},
  title        = {Using {GPT-5.2}},
  howpublished = {OpenAI API Documentation},
  year         = {2026},
  month        = jan,
  url          = {https://platform.openai.com/docs/guides/latest-model},
  note         = {Retrieved January 22, 2026, from the listed URL}
}

@misc{ref39_samsung_gear_sport_product,
  author       = {{Samsung}},
  title        = {Gear Sport ({SM-R600NZBAXAR})},
  howpublished = {Samsung},
  year         = {2026},
  month        = jan,
  url          = {https://www.samsung.com/us/mobile/wearables/smartwatches/gear-sport-blue-sm-r600nzbaxar/},
  note         = {Retrieved January 22, 2026, from the listed URL}
}

@misc{ref40_withings_sleep_analyzer_user_guide_v6,
  author       = {{Withings}},
  title        = {{Sleep Analyzer} v6.0},
  howpublished = {Withings Support},
  year         = {2023},
  month        = mar,
  url          = {https://support.withings.com/hc/article_attachments/13710141655825},
  note         = {[PDF]. Retrieved January 22, 2026, from the listed URL}
}

@misc{41_twilio_whatsapp_api,
  author       = {{Twilio}},
  title        = {Twilio API for {WhatsApp}},
  howpublished = {Twilio Documentation},
  year         = {2026},
  month        = jan,
  url          = {https://www.twilio.com/docs/whatsapp/api},
  note         = {Retrieved January 22, 2026, from the listed URL}
}

@misc{42_whatsapp_business_platform,
  author       = {{WhatsApp}},
  title        = {{WhatsApp Business Platform}},
  howpublished = {WhatsApp Business},
  year         = {2026},
  month        = jan,
  url          = {https://business.whatsapp.com/products/business-platform},
  note         = {Retrieved January 22, 2026, from the listed URL}
}

@misc{43_microsoft_functions_overview,
  author       = {{Microsoft}},
  title        = {What is {Azure Functions}?},
  howpublished = {Microsoft Learn},
  year         = {2025},
  month        = mar,
  url          = {https://learn.microsoft.com/en-us/azure/azure-functions/functions-overview},
  note         = {Page date: March 25, 2025. Retrieved January 22, 2026, from the listed URL}
}

\end{document}